\documentclass[usenatbib]{mn2e}
\usepackage{amsmath, amssymb}
\usepackage{graphicx}
\usepackage{cases}
\usepackage{aas_macros}

\title[The shape of a rapidly rotating n=1 polytrope]{The shape of a rapidly rotating polytrope with index unity}

\author[J.\ Knopik, P.\ Mach, A.\ Odrzywo{\l}ek]{Jerzy Knopik, Patryk Mach, Andrzej Odrzywo{\l}ek\\
Instytut Fizyki im.\ Mariana Smoluchowskiego, Uniwersytet Jagiello\'{n}ski, {\L}ojasiewicza 11, 30-348 Krak\'{o}w, Poland}

\begin{document}
\maketitle

\begin{abstract}
We show that the solutions obtained in the paper `An exact solution for arbitrarily rotating gaseous polytropes with index unity' by Kong, Zhang, and Schubert represent only approximate solutions of the free-boundary Euler--Poisson system of equations describing uniformly rotating, self-gravitating polytropes with index unity. We discuss the quality of such solutions as approximations to the rigidly rotating equilibrium polytropic configurations.
\end{abstract}


\section{Introduction}

In a recent paper \citet{kzs} attacked a long-standing problem of obtaining analytic solutions representing rigidly rotating stationary polytropes with the polytropic index $n = 1$ (polytropic exponent $\gamma = 1 + 1/n  = 2$). The quest for such solutions started with the works of \citet{chandraa}\footnote{This is actually the first of a series of four Chandrasekhar's papers on distorted polytropes \citep{chandraa,chandrab, chandrac, chandrad}.} and \citet{kopala, kopalb}, but the majority of papers on this subject has been published during three decades between 1960 and 1990 \citep{roberts, papoyan, blinnikov, hubbard, kozenko, cunningham, zharkov_book, caimmi, williams}. A modern account of these works can be found in \citet{horedt}.

\citet{kzs} assume that the shape of the boundary surface of such polytropes is an ellipsoid of revolution and construct analytic solutions of the corresponding equation for the distribution of the density. In this note we show that they satisfy only approximately the coupled system of Euler and Poisson equations describing equilibrium configurations of rigidly rotating polytropes with $n = 1$. We also clarify the difference between solving the free-boundary Euler--Poisson problem and the corresponding equation for the density that is solved in \citep{kzs}.

Solutions obtained by \citet{kzs} can still be understood as approximations to the rigidly rotating polytropes with $n = 1$, although the true shape of the boundary surface of such polytropes deviates from the ellipsoidal one. In general, the shape of an axisymmetric rotating polytropic body ($0<n<5$) can be described only approximately by a biaxial ellipsoid. The exact ellipsoidal description is possible only for the homogeneous, constant density polytrope $n=0$ \citep[cf.][pp.~82--84]{tassoul}.

In the following sections we discuss numerical solutions of rigidly rotating $n = 1$ polytropes, testing the ellipsoidal approximation. As expected, the deviation from the ellipsoidal shape is significant for rapidly rotating configurations. We also pay special attention to the model that was computed in \citet{kzs} assuming observational parameters of $\alpha$ Eri.

\section{Rotating polytropes with index unity}

Rigidly rotating axisymmetric equilibrium configurations of a self-gravitating perfect fluid are described by the Euler equations

\begin{subequations}
\label{euler}
\begin{align}
\frac{\partial P}{\partial \varpi} - \rho \Omega^2 \varpi & = - \rho \frac{\partial \Phi}{\partial \varpi}, \\
\frac{\partial P}{\partial z} & = - \rho \frac{\partial \Phi}{\partial z},
\end{align}
\end{subequations}
and the Poisson equation for the gravitational potential $\Phi$
\begin{equation}
\label{poisson}
\Delta \Phi = 4 \pi G \rho.
\end{equation}
Here $P$ denotes the pressure, $\rho$ is the mass-density, $\Omega$ is the angular velocity of the fluid, and $G$ is the gravitational constant. We work in cylindrical coordinates $(\varpi, \phi, z)$, assuming that the $z$ axis coincides with the rotation axis of the fluid.

For barotropic equations of state $P=P(\rho)$, one can introduce the specific enthalpy $h$ so that $dh = dP/\rho$. Then, in the region $U$ where $\rho > 0$, Eqs.\ (\ref{euler}) can be written as
\begin{equation}
\nabla (h + \Phi_c + \Phi) = 0,
\end{equation}
or, equivalently,
\begin{equation}
\label{euler_integrated}
h + \Phi_c + \Phi = C,
\end{equation}
where the centrifugal potential is
\begin{equation}
\label{aaf}
\Phi_c = - \frac{1}{2}\Omega^2 \varpi^2,
\end{equation}
and $C$ denotes an integration constant.

In this paper we deal exclusively with polytropic equations of state of the form $P = K \rho^{1 + 1/n}$, where $K$ is a constant, and assume $n = 1$. This yields $P = K \rho^2$, and a linear relation $h = 2 K \rho$. 

A rigidly rotating $n = 1$ polytrope is thus described by the Poisson equation (\ref{poisson}) and the equation
\begin{equation}
\label{aae1}
\nabla (2K \rho + \Phi_c + \Phi) = 0,
\end{equation}
that holds in the region $U$, and where $\Phi_c$ is given by Eq.\ (\ref{aaf}). It is important to stress that the boundary of the region $U$, at which the density $\rho$ tends to zero, is not known a priori, but has to be established by the procedure of solving Eqs.\ (\ref{poisson}) and (\ref{aae1}) itself. In mathematical literature, such a formulation is known as a free-boundary problem. It is a remarkable property of the set of Eqs.\ (\ref{poisson}) and (\ref{aae1}) that it allows for the determination of the boundary of $U$. In other words, Eqs.\ (\ref{poisson}) and (\ref{aae1}) do not admit regular solutions with a non-negative density $\rho$ for a wrong guess of the shape $U$.

The approximation presented by \citet{kzs} (and advertised as an exact solution) is based on the following, well known, observation. By computing the divergence of Eq.\ (\ref{aae1}) and combining the result with Eq.\ (\ref{poisson}), one obtains an equation for the density
\begin{equation}
\label{aad1}
\Delta \rho + \frac{2 \pi G}{K} \rho = \frac{\Omega^2}{K}.
\end{equation}
The above equation also holds in the region $U$, and it is assumed that $\rho$ tends to zero at the boundary of $U$. Equation (\ref{aad1}) is equivalent to Eq.\ (8) of \citet{kzs}, which is written in oblate spheroidal coordinates, and in fact it is Eq.\ (\ref{aad1}) and not the system of Eqs.\ (\ref{poisson}) and (\ref{aae1}) that is solved by \citet{kzs}. In contrast to the system of Eqs.\ (\ref{poisson}) and (\ref{aae1}), it admits solutions with the boundary of $U$ being an ellipsoid of revolution.

It is clear from the above derivation that solutions of Eq.\ (\ref{aad1}) and the distribution of the density $\rho$ obtained by solving Eqs.\ (\ref{poisson}) and (\ref{aae1}) may differ by a function $f$ satisfying $\Delta f = 0$. It is also easy to see that Eq.\ (\ref{aad1}) yields the proper solution, i.e. a solution equal to that of Eqs.\ (\ref{poisson}) and (\ref{aae1}), provided that the true shape of the boundary of $U$ is assumed. The fact that this boundary cannot be an ellipsoid of revolution follows from the classic theorem that can be found in the textbook by \citet[][pp.\ 82--84]{tassoul}. It states that there are no rigidly rotating, barotropic solutions of Eqs.\ (\ref{euler}) and (\ref{poisson}) with an ellipsoidal boundary, except for the configurations of constant density.\footnote{The theorem is actually more general: it excludes all barotropic configurations with an ellipsoidal stratification.}


\section{Numerical solutions}
\label{sec_numerical}

\begin{figure*}
\includegraphics[width=\linewidth]{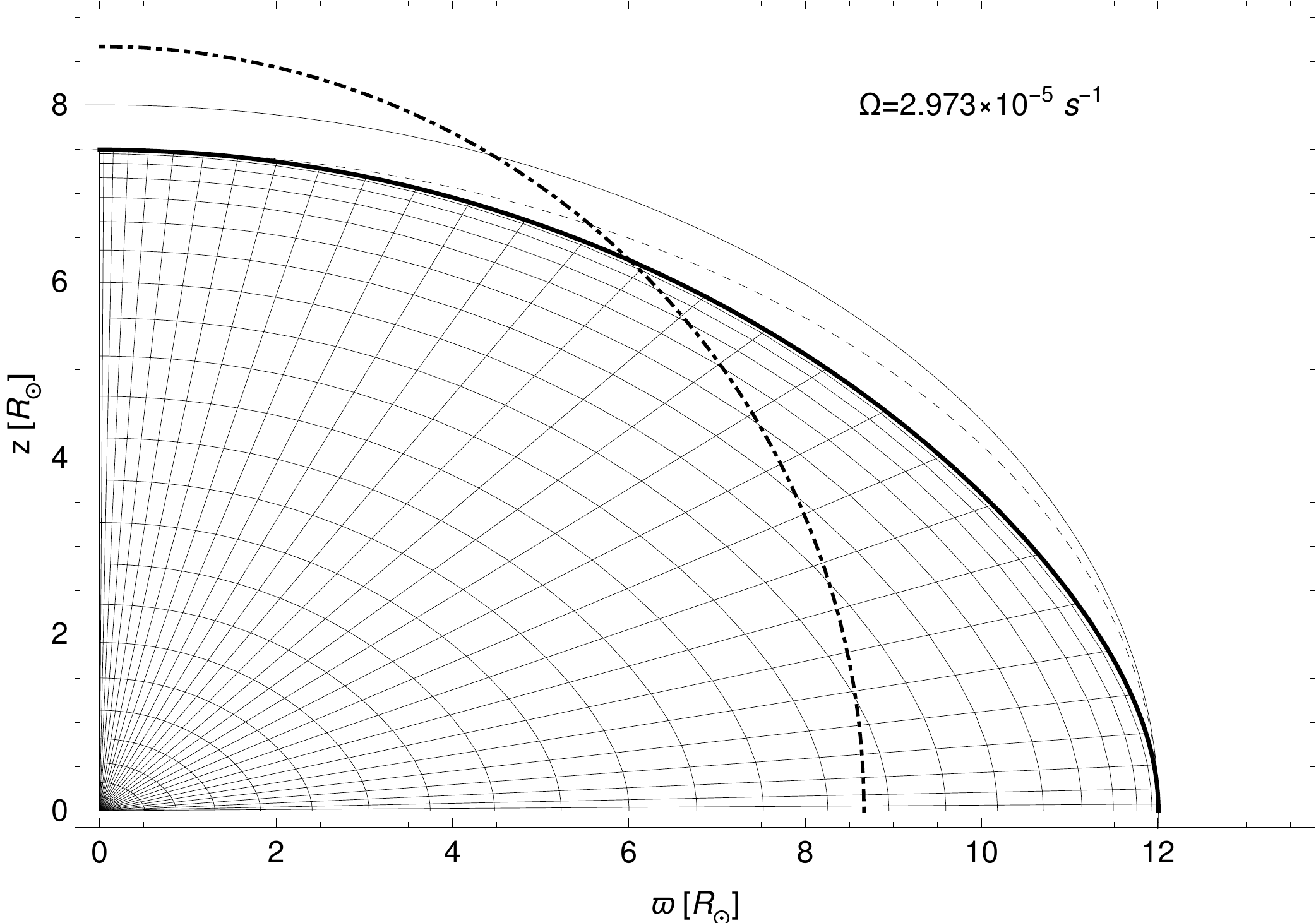}
\caption{\label{fig1} Meridional cross-section of a rigidly rotating $n=1$ polytrope obtained numerically for the model of $\alpha$ Eri, as explained in the text. The thick solid line shows the computed surface, while the thin one is a part of an ellipse with semi-axes that are identical to the polar and equatorial radii of the \citet{kzs}, i.e., $R_p/R_e = \sqrt{1-\mathcal{E}^2} = 2/3$. The thin dashed line depicts an ellipse with the semi-axes equal to the values of $R_p$ and $R_e$ that were obtained for our numerical model of $\alpha$ Eri. The thick dot-dashed line depicts the shape of a non-rotating configuration with the same mass and the same value of the polytropic constant $K$.}
\end{figure*}

\begin{figure*}
\includegraphics[width=\linewidth]{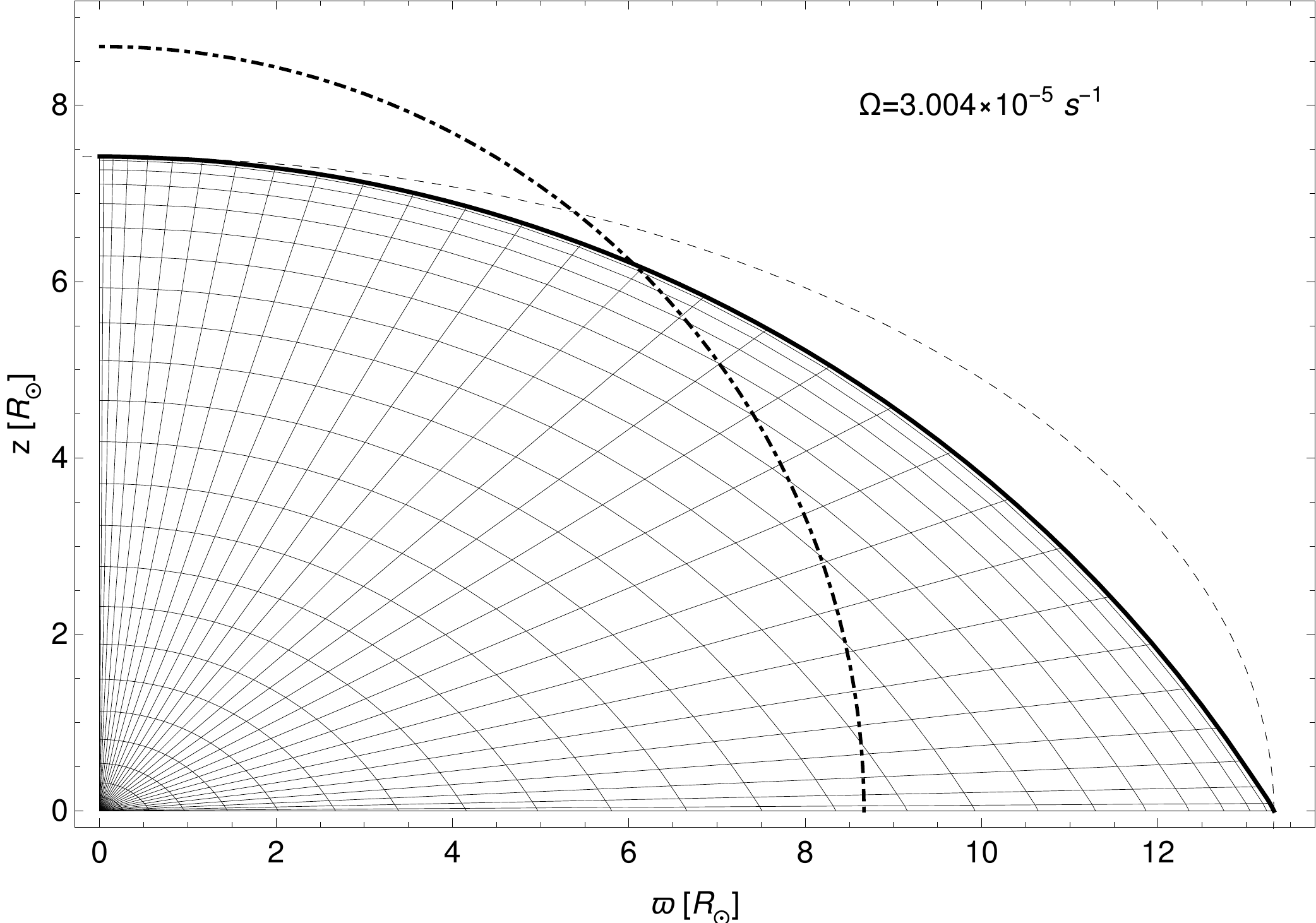}
\caption{\label{fig2} Same as Fig.\ \ref{fig1}, but for the model of an almost critically rotating polytrope. Here $\Omega= 3.004 \times 10^{-5} \, \text{s}^{-1}$.}
\end{figure*}

\begin{figure*}
\includegraphics[width=\linewidth]{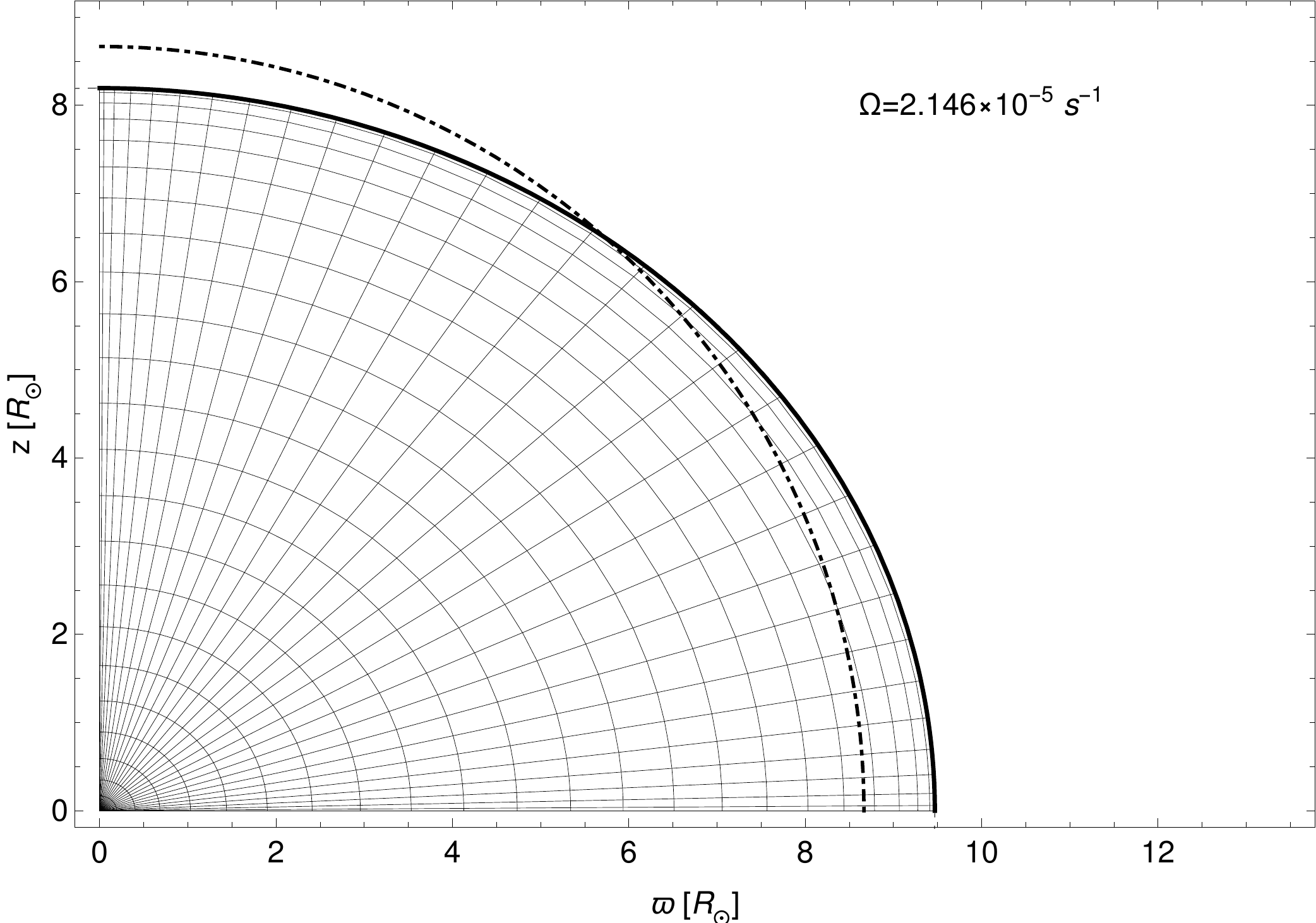}
\caption{\label{fig3} Same as Fig.\ \ref{fig1}, but for the model of a slowly rotating polytrope with $\Omega = 2.146 \times 10^{-5} \text{s}^{-1}$. The difference between the obtained shape of the boundary surface and the ellipsoidal fit is not visible in this case.}
\end{figure*}

Because guessing of the right shape of the region $U$ in Eq.\ (\ref{aae1}) is difficult (if at all possible), one has to resort to numerical methods or a combination of numerical and analytic schemes \citep[see, e.g.,][]{ao}. 

In this section we give numerical examples of solutions of Eqs.\ (\ref{poisson}) and (\ref{aae1}), obtained without assuming that the surface of the configuration is ellipsoidal. Our algorithm follows strictly the scheme proposed by \citet{eriguchi_mueller}. In short, it is based on Eq.\ (\ref{euler_integrated}), which is discretised on a two-dimensional grid. The values of the gravitational potential $\Phi$ at the grid nodes are computed using the expansion of the Green function for the Laplace operator in terms of Legendre polynomials. In this way Eqs.\ (\ref{poisson}) and (\ref{aae1}) are converted into a system of algebraic equations, which is then solved using standard numerical methods.

The method of \citet{eriguchi_mueller} allows one to obtain whole sequences of models, say with a fixed value of the total mass and different values of the angular momentum. It is important to stress that due to the scaling symmetry of Eqs.\ (\ref{poisson}) and (\ref{aae1}), only one such sequence needs to be computed \citep[see, e.g.,][]{aksenovblinnikov}.

In Figs.\ \ref{fig1}--\ref{fig3} we plot the shapes of three of the obtained configurations. Because \citet{kzs} discuss a model of $\alpha$ Eri with an ellipsoidal surface, we believe that a comparison of the numerical solution and their analytic approximation for the observed parameters of $\alpha$ Eri would be in order. Such a comparison is shown in the first of the presented figures.

Our numerical solution from Fig.~\ref{fig1} was computed with the observed mass $M = 4.9 M_\odot$, the equatorial radius $R_e = 12.0 R_\odot$, and the angular velocity $\Omega = 2.9725 \times 10^{-5} \mathrm{s}^{-1}$, as quoted by \citet[][Table 1]{kzs}. The values of solar parameters used in the code are: $M_\odot = 1.98855 \times 10^{33}$~g, $R_\odot = 6.95700 \times 10^{10}$~cm. The assumed value of the gravitational constant reads in cgs units $G = 6.67384 \times 10^{-8} \, \mathrm{cm^3 s^{-2} g^{-1}}$.

Our computed ratio of polar and equatorial radii for the $\alpha$ Eri model is Rp/Re = 0.624, and the polytropic constant
$K= 1.54 \times 10^{16} \, \mathrm{cm}^5 \, \mathrm{g}^{-1} \, \mathrm{s}^{-2}$. \citet{kzs} report the value $K = 1.7500 \times 10^{16} \, \mathrm{cm}^5 \, \mathrm{g}^{-1} \, \mathrm{s}^{-2}$ and the value of the eccentricity $\mathcal{E} = 0.745356$, which, up to the used precision, is equivalent to $R_p/R_e = \sqrt{1 - \mathcal{E}^2} = 2/3$.
In our computation the value of $K$ is obtained together with the structure of
the configuration as one of the unknowns \citep{eriguchi_mueller}. The obtained value differs by 12\% from that of \cite{kzs} because of the differences in the density stratification. The obtained shape of the boundary surface is plotted in Fig.\ \ref{fig1}. For comparison, we also plot an ellipse with the semi-axes that were obtained by \citet{kzs} and an ellipse fitted to the obtained shape, i.e., an ellipse with semi-axes equal to the values of $R_p$ and $R_e$ of our numerical model. There is a clear discrepancy between all these shapes.

The solution presented in Fig.~\ref{fig1} here is relatively close to the limiting configuration characterized by the maximal allowed (critical) rotation. Finding the ratio of $R_p/R_e$ corresponding to the critical configuration is a subtle numerical issue. \citet[][Table 4]{james} and \citet[][Table 1]{hurley} give the values 0.558 and 0.493, respectively. \citet[][Table 7]{cook}, who consider general-relativistic configurations report a Newtonian-limit polytropic solution with $n = 1$ and $R_p/R_e = 0.552$. The shape of an (almost) critically rotating $n = 1$ polytrope found in our numerical computations is shown in Fig.\ \ref{fig2}. In this example we keep the same values of the total mass $M = 4.90 M_\odot$ and the polytropic constant $K$ as before. The remaining parameters of this solution are: $R_p = 7.42 R_\odot$, $R_e = 13.3 R_\odot$, $R_p/R_e = 0.557$, $\Omega = 3.004 \times 10^{-5} \, \text{s}^{-1}$.

As expected, the ellipsoidal approximation is much better for slowly rotating polytropes. A comparison between the actual shape and the matching ellipsoid for $\Omega =  0.2146 \times 10^{-5} \, \text{s}^{-1}$ is shown in Fig.\ \ref{fig3}. Here again we keep the same values of $M$ and $K$. The remaining parameters read: $R_p = 8.19 R_\odot$, $R_e = 9.47 R_\odot$, $R_p/R_e = 0.865$. Note that all solutions presented in Figs.\ \ref{fig1}--\ref{fig3} form a sequence characterized by the same mass and the same equation of state, but different angular velocities.

The numerical grid used in our computations is also depicted in each of Figs.~\ref{fig1}-\ref{fig3}. We use 31 grid points in the angular direction and 23 in the radial one. The highest order of Legendre polynomials used in the numerical code is 40.  

\section{Final remarks}

Polytropic solutions with index unity serve as suitable approximations of the structure of Jovian planets \citep{hubbard}. In a recent work \citet{cao} provide an interesting comparison between parameters of a polytropic model and the data for Jupiter. [Note that it is explicitly stated in \citet{cao} that the outer boundary in the polytropic model differs from an exact ellipsoid of revolution.] The values of $n$ close to unity can be probably also assumed for brown dwarfs and some neutron stars. For the star in the radiative equilibrium (like $\alpha$ Eri) the value $n \approx 3$ seems to be more appropriate. This fact was actually admitted by \citet{kzs}.

It is possible to search for exact solutions representing rotating polytropes of index $n = 1$ assuming an ellipsoidal boundary surface and differential rotation. Such solutions were constructed by \citet{cunningham} (He uses the term `spheroids' referring to ellipsoids of revolution). 

It is also worth pointing out that because every solution of Eqs.\ (\ref{poisson}) and (\ref{aae1}) satisfies Eq.\ (\ref{aad1}), the latter can be actually used to infer important characteristics of solutions of Eqs.\ (\ref{poisson}) and (\ref{aae1}). This is also true in nonlinear cases (e.g., for polytropes with $n \neq 1$); results of this type were obtained by \citet{sobolev}.

Stability of the rotating $n = 1$ polytropes was investigated by \citet{bini}.

\section*{Acknowledgements}

We are grateful to Georg Horedt for his comments on the first version of this article. A.O. was supported by the Kosciuszko Foundation.

\bibliographystyle{mn2e}
\bibliography{comment4}

\end{document}